\documentclass[12pt,draft]{IEEEtran} 

\usepackage{amsfonts}
\def\Btl{\tilde{ B}\,}

\def\eb{{\bf e}}

\def\Ib{{\mathbb{I}}\,}
\def\Dc{{\cal{D}}\,}

\def\Ah{{\hat{A}}\,}
\def\Bh{{\hat{B}}\,}

\def\kh{{\hat{\kappa}}\,}
\def\tb{{\overline{\tau}}\,}
\def\lb{{\overline{\lambda}}\,}
\def\vb{{\bf v}}
\def\wb{{\bf w}}
\def\zb{{\overline{z}}\,}

\def\limfunc#1{\mathop{\rm #1}}
\def\ni{\noindent}

\def\eopp{{ \vrule height7pt width7pt depth0pt} }
\def\Prf{\noindent {\em Proof:}\, }
\def\proof{\noindent {\em Proof:}\, }

\def\eopp{{ \vrule height7pt width7pt depth0pt} }
\def\Prf{\noindent {\em Proof:}\, }

\makeatletter
\@addtoreset{equation}{section}
\makeatother
\newtheorem{thm}{Theorem}[section]
\newtheorem{defn}[thm]{Definition} 
\newtheorem{lem}[thm]{Lemma} 
\newtheorem{cor}[thm]{Corollary} 
\newcommand{\BEQ}{\begin{equation}}
\newcommand{\EEQ}{\end{equation}}
\newcommand{\NEQ}{\end{equation}}
\begin{document}
\bibliographystyle{plain}
\title{%
EXPONENTIAL CONDITION NUMBER OF
SOLUTIONS OF THE DISCRETE LYAPUNOV EQUATION}

\author{
Andrew P.~Mullhaupt$^*$ and Kurt S.~Riedel${^\dagger}$
\thanks{* S.A.C. Capital Management, LLC,
540 Madison Ave, New York, NY 10022, \ \
$\dagger$  Millennium Partners, 666 Fifth Ave, New York 10103 \ \
The authors thank the referee for his detailed comments.}
}

\maketitle

\begin{abstract}
The condition number of the $n\times n$ matrix $P$ is examined, where
$P$ solves 
$P-APA^*= BB^*$,
and $B$ is a $n \times d$ matrix.
Lower bounds on the condition number, $\kappa$, of $P$ are given
when $A$ is normal, a single Jordan block or in Frobenius form.
The bounds show that the ill-conditioning of $P$ grows
as $\exp(n/d) >> 1$.
These bounds are related to the condition number of the transformation
that takes $A$ to input normal form.
A simulation shows that $P$ is typically ill-conditioned in the case of
 $n>>1$ and $d=1$. When $A_{ij}$ has an independent Gaussian distribution
(subject to restrictions),
we observe that $\kappa(P) ^{1/n} \sim 3.3$. The effect of autocorrelated
forcing on the conditioning on state space systems is examined.
\end{abstract}


EDICS Numbers: 2-PERF 2-IDEN, 2-SYSM


\newpage
\section{INTRODUCTION}
\label{I}
In system identification, one needs to solve linear algebraic systems:
$P{c}= {f}$, where ${c}$ and  ${f}$ are $n$-vectors
and $P$ is the controllability Grammian, i.e.\ $P$ is
the $n\times n$ positive definite matrix that solves
\BEQ \label{SteinEq}
P-APA^*= BB^* \ .
\NEQ
Equation (\ref{SteinEq}) is known as the
discrete Lyapunov equation  and is more properly
called  Stein's equation.
In (\ref{SteinEq}), the  $n\times n$ matrix $A$ and the $n\times d$ matrix $B$
are given. The matrix $A$ is known as the state advance matrix and the
matrix $B$ is known as the input matrix.
Together, $(A,B)$ is known as an input pair.
We assume that $A$ is stable
and that $(A,B)$ is controllable.
In this case, there is an unique selfadjoint solution of
(\ref{SteinEq}) and it is positive definite \cite{LT}.
We denote the solution of (\ref{SteinEq}) as a function
of $A$ and $B$ by $P(A,B)$

We study the condition number of $P(A,B)$,  $\kappa(P)
 \equiv \kappa(P(A,B)) \equiv \sigma_1(P(A,B)) /$ $ \sigma_n(P(A,B))$,
where $\sigma_1(P)$ is the largest singular value of $P$
and $\sigma_n(P)$ is the smallest.
We consider cases where the system input dimension, $d$, is smaller than the
state space dimension, $n$. In this case, we claim that the condition number
of $P$, $\kappa(P)$ can be exponentially large in $n/d$.
Since the case $n>>d$ is common in signal processing and
system identification, our results put strong limitations
on the applicability of  high order
arbitrary state space realizations.

A number of bounds  on either $\sigma_1(P(A,B))$  or $\sigma_n(P(A,B))$
exist in the literature \cite{Mori,Komaroff,Kwon,GLaub,Lee,Tippett}.
Many of these bounds require that $\det(BB^*)>0$ to be nontrivial.
Theorem 2.2 of \cite{Tippett} can be used to bound the ratio
of $\sigma_1(P)/\sigma_n(P)$. (See also \cite{Tippett2}.)
The existing bounds on $\sigma_i(P(A,B))$ generally make no assumptions
on $(A,B)$ and therefore tend to be weak or hard to evaluate.
If $A$ is real, symmetric, and stable, Penzl \cite{PenzlBnd} gives a bound which
we describe in Section \ref{ADISect}.
For the continuous time case, interesting results on the condition number may be
found in \cite{ASZ}

Our lower bounds on  $\kappa(P(A,B))$ are for specific, commonly considered
classes of input pairs, $(A, B)$, such as {\em companion} matrices and
{\em normal} matrices and when $A$ is a single Jordan block.

Our results are based on transforming the input pair, $(A,B)$, into 
input normal (IN) form. Input normal form  implies that
the identity matrix solves the
discrete Lyapunov equation. 
Input normal pairs have special representations that allow for fast matrix-vector operations when $A$ is a Hessenberg matrix \cite{MR1,MR2,MRbook}.

In \cite{NH}, a numerical simulation  shows that input
normal filters perform well in the presence of autocorrelated noise.
We examine the condition number of the controllability Grammian
when forcing term is autocorrelated. We derive a bound that explains
the good performance of IN filters \cite{NH}.
Other advantages of IN filters are described in \cite{MRbook,NHG,Ninness}

The condition number of $P$ is related to two other well-known problems:
a) the distance of an input pair to the set of uncontrollable pairs
\cite{Eising}
and b) the sensitivity of the $P$ to perturbations in $(A,B)$
\cite{GahLaub,GLaub}.
It is well known that $1/\kappa(P)= \min\{ \|E\|_2 /\|P \|_2: (P+E)\
 {\rm is\ singular}\}$ \cite{Higham}.
Thus we can lower bound the distance to uncontrollability
by the $1/\kappa(P)$ times the sensitivity of the
discrete Lyapunov equation. 
Our results indicate that $1/\kappa(P)$ is typically exponentially small in $n/d$.

We present numerical simulations which compute the distribution of
$\kappa(P(A,B))$ for several classes of input pairs, $(A, B)$.
When  the elements of $(\sqrt{n+2}A,B)$ are independently distributed
as Gaussians with unit variance, our
simulation  shows that the ensemble average of
 $\kappa(P)^{1/n}$ tends to a constant for $d=1$.
We observe that $\log(\log(\kappa(P)))$ is approximately Gaussian.
These numerical results indicate that the ill-conditioning
problems of  $\kappa(P)$ are probably generic when $n/d <<1$.
To accurately solve (\ref{SteinEq}), we use a novel
{\em QR iteration to precondition (\ref{SteinEq})
and then apply a square root version of the doubling method} \cite{Benner}.

Section \ref{GenSimSect} presents our computation of $\kappa(P(A,B))$
for an ensemble of stable controllable input pairs.
Section \ref{INSect} defines IN form and present new results on the properties of IN pairs.
Section \ref{TransfSect} gives lower bounds on the condition number based
on the transformation to an IN matrix and
applies the bound to the case when $A$ is normal.
Section \ref{ADISect} gives abstract bounds based on the ADI iteration.
Section \ref{CompSect} gives lower bounds when $A$ is in {companion form}.
Sections \ref{PowerSect} and \ref{FracSect} give additional bounds for
normal $A$.
Section \ref{JordanSect} gives bounds when $A$ is a single Jordan block.
Section \ref{ColSect} examines condition numbers of the state covariance when the
system is forced with colored noise.

{\em Notation:} 
Here $A$ is a $n\times n$ matrix
with eigenvalues $\{\lambda_i\}$ ordered as
 $1 \ge |\lambda_1 | \ge |\lambda_2 | \ldots \ge  |\lambda_n|$ and
singular values,  $\sigma_1 \ge \sigma_2\ge  \ldots \ge  \sigma_n$.
Depending on context, $\Lambda$ is the n-vector of $\lambda_i$
or the corresponding diagonal matrix.
The matrix $B$ has dimension $n\times d$.
When $A$ is stable and $(A,B)$ is controllable,
we say that the input pair $(A,B)$ is CS.
If $A$ is also invertible, we say  $(A,B)$ is CIS.
For us, `stable'' means $|\lambda_1 |<1$,
sometimes known as strict exponential stability. We let
$\Dc(A,B)$ denote the set of stable, controllable $(A,B)$ input pairs of dimension
$ n\times n$ and  $n\times d$. The  $ n\times n$ identity matrix is denoted
${\mathbb{I}}_n$.
The transformation matrices, denoted by $T$ and $U$,
have  dimension $ r \times n$  and rank $r$, where $r \le n$.
The Moore-Penrose inverse of $T$ is denoted by $T^+$.
Here $\| \cdot \|_2$ is the Frobenius norm while $\| \cdot \|$
is any unitarily invariant matrix norm


\noindent
\section{GENERIC CONDITION NUMBER} \label{GenSimSect}
We begin by examining the probability distribution of condition number
of $P(A,B)$ as $A$ and $B$ are varied over a probability distribution,
 $\nu(A,B)$, on stable, controllable input pairs.
 We limit ourselves to single input pairs ($d=1$).

A common class of random matrices is
 $\{A | A_{ij} \sim N(0, 1) / \sqrt{n} \}$,
with the probability measure $\mu(A)$.
The Girko law \cite{Edelman} states that
the eigenvalues of such random $A$ 
are  uniformly distributed
on the complex disk $|\lambda_i| <1$ as $n \rightarrow \infty$.
For finite $n$, the distribution of eigenvalues is given in
Theorem 6.2 of \cite{Edelman}.
We exclude unstable $A$ and uncontrollable $(A,B)$ in our studies.
We normalize $A$ by $1/\sqrt{n+2}$ instead of $1/\sqrt{n}$ to improve
the odds of obtaining stable $A$.
Specifically, we define the distribution:

\begin{defn} \label{DefDAB1}
Let $\nu(A,B)$ be the probability measure induced on $\Dc(A,B)$ by letting
the matrix elements $ A_{ij}\sqrt{(n+2)}$  and $B_{ij}$ have independent
Gaussian distributions, $N(0, 1)$, subject to the CS restriction.
\end{defn}

Each probability distribution on $(A,B)$ induces a distribution of $P(A,B)$
and $\kappa(P(A,B))$.
We simulate the induced distribution by solving the
discrete Lyapunov equation 
for 2,500 $(A,B)$ pairs chosen from the distribution $\nu$.

{\em Inaccurate numerics will tend to underestimate $\kappa(P)$.}
Even for $n \approx 10$, these systems can be so ill-conditioned that
existing numerical methods inaccurately determine the condition number.
Therefore, we developed new numerical algorithms for the solution
of  
(\ref{SteinEq}) \cite{MR4}. To solve the
discrete Lyapunov equation, 
we use a novel square root version of the doubling method.
For ill-conditioned problems, we find that {\em preconditioning the
discrete Lyapunov equation 
is important to accurately evaluate  
 the condition number of $P$}  \cite{MR4}.



Table 1 gives the quantile distribution of $\log (\kappa(P))$
as a function of $n$ for our numerical simulation with
 $(A,B)$ distributed in $\nu(A,B)$. 
 {\em The median condition number scales as  $\log (\kappa(P)) \approx 1.2 n$.
The interquartile distance is approximately independent of $n$
with a value of $\approx 4.4$.} (The interquartile distance is the distance between the
$75$th percentile and the $25$th percentile and is a measure of the width of the distribution.)
If the distribution were normal, the interquartile distance would
be roughly 1.35 standard deviations.
We plotted the quantiles of $\log(\log(\kappa(P)))$ and of $\log(\kappa(P))$
versus the quantiles of the Gaussian distribution.
{\em These quantile-quantile plots 
show that $\log(\log(\kappa(P)))$ has an approximately
Gaussian distribution}
and that  $\log(\kappa(P))$ has wide tails.
Naturally, the tails of the empirical distribution are more poorly determined
than the median and the quartiles.

\medskip
\begin{center}
\begin{tabular}{|c|c|c|c|c|c|c|c|}
\hline
n  & 1\% & 10\%  & 25 \% & 50 \% & 75\%  & 90\% & 99 \% \\
\hline
8 &  6.23  & 8.22  & 9.62  & 11.3  & 13.4  & 15.8  & 20.2  \\
16& 14.2  & 16.5  & 18.2  & 20.3  & 22.5  & 24.9  & 30.4  \\
24& 22.3  & 25  & 26.7  & 28.7  & 31.1  & 33.6  & 38  \\
\hline
\end{tabular}
\end{center}

 Table 1: Quantile distribution of $\log (\kappa(P))$
for $(A,B)$ distributed in $\nu(A,B)$ as a function of $n$.

In \cite{Edelman1}, it is shown that a random matrix, $A$, has
${\rm E}[ \log(\kappa(A))] \sim \log(n)$, where $E$ denotes the expected value. 
Thus $P(A,B)$ is typically much more
poorly conditioned than $A$ is.
In \cite{VT}, it is shown that a random lower triangular matrix, $L$, has
 $\kappa(L)^{1/n} \approx 2$ with probability tending to 1
as $n \rightarrow \infty$.
For the median value of $\kappa(P)$ in our computation,
{\em the Cholesky factor of $P$, $L$ scales as
 $\kappa(L)^{1/n} \approx 1.8$}, which is nearly as badly conditioned as
those in \cite{VT}.


Table 1 displays results for $d=1$. Empirically, we observe that
the condition number grows at least as fast as $n/d$.
In Section \ref{TransfSect}, we derive  lower bound for the condition number
when $A$ is normal. We apply this bound to each matrix in our simulation.
Table 3 shows that the actual condition number is worse
than the normal bound by a factor of roughly 100 on average.

\noindent
\section{INPUT NORMAL PAIRS} \label{INSect}
In examining the condition number of solutions of the
discrete Lyapunov equation, 
it is natural to begin with input pairs that admit solutions
with condition number one.

\begin{defn} \label{DefTIN}\label{TINdef}
A  input pair, $(\tilde{ A},\tilde{ B})$, is
input normal  (IN) of grade $d$ if and only if
$\tilde{A}$ is stable, $\limfunc{ rank} (\Btl) = \limfunc{column\ dim}(\Btl) =d$, and
\BEQ \label{E7}\label{INeq}
\tilde{ A}\tilde{ A}^{*} =  { \Ib} -
\tilde{ B}\tilde{ B}^{*} \ \ \ .
\EEQ
A matrix, $\tilde{ A}$, is a IN
 matrix of grade $d$ if and only if
there exists a $n \times d$-matrix $\tilde{B}$ such that
$(\tilde{ A},\tilde{ B})$, is an
IN pair.
If  $\tilde{ A}$ is lower (upper)
triangular as well, $(\tilde{ A},\tilde{ B})$ is a triangular input normal pair.
If  $\tilde{ A}$ is
Hessenberg as well, $(\tilde{ A},\tilde{ B})$ is a
Hessenberg input normal pair.
\end{defn}

In \cite{MRbook}, `input normal pairs'' are called orthogonal filters.
In \cite{Moore}, `input normal'' has a more restrictive
definition of (\ref{INeq}) and the additional requirement that the
observability Grammian is diagonal.
In our definition of `input normal'',
we do not impose any such condition on the observability Grammian.
We choose this language so that `normal'' denotes restrictions on only
one Grammian while `balanced'' denotes simultaneous restrictions on
both Grammians \cite{Laubetal,Moore}.
This usage is consistent with the definitions of \cite{Dewilde}.
{\em Input normal $A$ are generally not normal matrices.}

By Theorem 2.1 of \cite{AM},
if the controllability Grammian is positive definite,
then the input pair is stable.
In \cite{Ober}, Ober shows that stability plus a positive definite
solution to the 
discrete Lyapunov equation, 
(\ref{SteinEq}), implies that the input pair is controllable.
Thus for IN pairs, stability is equivalent to controllability.
We now show that any CS input pair may be transformed to an IN pair.

\begin{thm}
\label{MINthm} \cite{MRbook}
Every  stable,  controllable input pair
$(A, B)$, is similar
to a input normal pair
$(\tilde{ A} \equiv { TAT}^{-1}, \tilde{ B}\equiv { TB})$
with $\|\tilde{B}\|^2 \le 1$.
\end{thm}

\Prf The unique solution of (\ref{SteinEq}), ${ P}$,
is strictly positive definite \cite{LT}.
Let $L$ be the  unique Cholesky lower triangular factor of ${ P}$
with positive diagonal entries, ${ P} ={ LL}^{*}$.
We set  $T=L^{-1}$, $\hat{ A}={ L}^{-1}{ AL}$, and $\hat{ B}={ L}^{-1}{ B}$.
\eopp

Using the singular value decomposition, we have the following characterization
of IN matrices:

\begin{thm} \label{INconstruct} \label{SVDchar}
Let $A$ be a stable $n\times n$ matrix with $\sigma _{1}\left( A\right) =1$,
and  let $d$ equal the number of singular values  of $A$ less than 1,
($d=\#\left\{ k|\sigma _{k}\left( A\right) <1\right\}$).
There is an $n\times d$ matrix $B$ with $\limfunc{rank}(B)=d$ such that
${\mathbb{I}}-AA^{*}=BB^{*}$ and therefore $A$ is an IN matrix.
The smallest $d$ singular values of $A$ satisfy 
$\prod_{j=n-d+1}^n \sigma_j^2(A) =\prod_{i=1}^n |\lambda_i|^2$, 
where $\lambda_i$ are the eigenvalues of $A$.
\end{thm}


\proof
Let $v_k$ be the $k$th singular vector of ${\mathbb{I}}-AA^{*}$ and define
$B= (\sqrt{1-\sigma_n(AA^*)}v_{n},$ $\sqrt{1-\sigma_{n-1}(AA^*)} v_{n-1}, \ldots 
\sqrt{1-\sigma_{n-d+1}(AA^*)}v_{n-d+1})$.
This constructs $B$.
The singular value identity follows from
$\prod_{j=n-d+1}^n \sigma_j^2 = \prod_{j=1}^n \sigma_j^2(A) =\det(AA^*)=
|\det(A)|^2=\prod_{i=1}^n |\lambda_i|^2$.
\eopp

For input normal pairs,
this yields the bound: 
$\sigma_n^2(A) \le \left(\prod_{j=n-d+1}^n \sigma_j^2(A)\right)^{1/d}$
$= \left(\prod_{1}^n |\lambda_j|^2\right)^{1/d}$.

There are many similar input normal pairs 
since if $(A, B)$ is IN, then so is
$(UAU^*, UB)$ for any orthogonal $U$.
This additional freedom may be used to simplify
the input pair representation \cite{MRbook,MR2,MR3}.

\ni
\section{CONDITION NUMBER BOUNDS AND THE TRANSFORMATION TO INPUT NORMAL PAIRS}
\label{TransfSect}

In this section, we derive lower bounds on the condition number of $P(A,B)$.
Our bounds are based on transforming $(A, B)$ to an IN pair $(A', B')$.
The following lemma describes the
transformation of solutions under a linear change of coordinates.

\begin{lem} \label{AbstractBnd}
Let $T$ be an $r \times n$ matrix of rank-$r$ with $r \le n$ and let
the rows of $T$ be a basis for a left-invariant subspace of $A$.
Define $A'$ by  $TA= A'T$. Let $\| \cdot \|$ be an unitarily invariant 
matrix norm
and let $\phi$ be an analytic function on
the spectrum of $A$ with ${\left\| \phi(A)\right\| }>0$. Then
$ \kappa(T) \equiv \sigma_1(T) \sigma_1(T^{+})  \ge  
{\left\| \phi(A')\right\| } \left/
{\left\| \phi(A)\right\| }\right.$.
When $A$ is invertible 
$ \kappa(T) 
\ge  {\left\| A'^{-1}\right\| } \left/  {\left\|A^{-1} \right\|} \right.$.
Also $\|T \| \|T^{+} \| \ge \kappa(T)  \|\eb_1\eb_1^*\|^2$,
where $\eb_1$ is the unit vector in the first coordinate.
\end{lem}

\Prf Note $\phi(A') = T\phi(A)T^{+}$ since $TT^{+}=\Ib_r$.
We apply the bound $\|FGH \| \le \sigma_1(F) \sigma_1(H) \|G \|$
\cite[p.\ 211]{HJ2} to $\phi(A') = T\phi(A)T^{+}$.
When $A$ is invertible, so is $A'$ and $A'^{-1} = TA^{-1}T^{+}$.
To bound  $\|T \| \|T^{+} \|$, we use the bound $\|T \|> \sigma_1(T)\|\eb_1\eb_1^*\|$
\cite[p.\ 206]{HJ2}.
\eopp

 A related result in \cite[p.\ 162]{HJ2}  is
\BEQ
\kappa(T) \ge \max\{\sigma_k(A) / \sigma_k(A') ,\ \sigma_k(A') / \sigma_k(A) \} \ ,
\NEQ
for invertible $T$ and nonvanishing $\sigma_k(A')$ and $\sigma_k(A)$.

When $r=n$, $T$ is invertible and $A'$ is similar to $A$: $A'=TAT^{-1}$.
In this case ($r=n$), we can reverse the roles of $A$ and $A'$ in the bounds as well.
The case $r<n$ is of interest in model reduction problems,
where one projects a system onto a left invariant subspace of $A$.

In the remainder of this section, we use $\phi(A)=A$ and $\phi(A)=A^{-1}$.
When $A'= TAT^{-1}$ is input normal,
we have the following bound for the condition number of the transformation of
a stable matrix $A$ to input normal form.

\begin{thm} \label{TransBnd}
Let $A$ be stable and invertible and $A'\equiv TAT^{-1}$ be an input normal
matrix of grade $d$, where
$T$ is an invertible matrix and $d<n$, then
\BEQ \label{TbndEq}
 \kappa_{}(T) 
\ge \max \left\{ \sigma_1\left(A\right) , \
\frac{1 }{\sigma_1\left( A\right) },\
\frac{\sigma_n\left(A \right) }{ \prod_{i=1}^n |\lambda_i(A)|^{1/d}}, \
\frac{\sigma_n\left( A'\right) }{\sigma_n\left( A\right) }
\right\} \ ,
\NEQ
where $\{ \lambda_i(A)\} $ are the eigenvalues of $A$
and $\sigma_1(A)$ and $\sigma_n(A)$ are the
largest and smallest singular values of $A$. For $d=1$,
$\sigma_n(A')= \prod_{i=1}^n |\lambda_i(A)|$.
\end{thm}

{\em Proof:} By Theorem  \ref{SVDchar}, $\sigma_1(A')=1$ and
$\sigma_n(A')\le \prod_{j=n-d+1}^n \sigma_j^{1/d}(A') = |{\det}(A'A'^*)|^{1/2d}
= |{\det}(A)|^{1/d} = \prod_{i=1}^n |\lambda_i(A)|^{1/d}$.
\eopp

Note that Theorem \ref{TransBnd} does not depend any specific input matrix $B$.

\begin{cor} \label{PbndCor}
Let $(A, B)$ be a CIS  
input pair, then the condition number of $P(A,B)$ satisfies the equality $\kappa_{}(P(A,B))= \kappa_{}(T)^2$,
where $\kappa_{}(T)$ and $A'$ are defined in
Theorem \ref{TransBnd}. 
\end{cor}

\Prf
The unique solution of (\ref{SteinEq}), ${ P}$,
is strictly positive definite \cite{LT}.
Let $L$ be the Cholesky factor of $P(A,B)$: $LL^*=P(A,B)$, and set $T=L^{-1}$.
Note $\kappa_{}(P(A,B)) = \kappa_{}(T^{-1}T^{-*})=\kappa_{}(T)^2$.
\eopp

For normal advance matrices, $\sigma_n\left(A \right) = |\lambda_n(A)|$,
the smallest eigenvalue of $A$. This simplifies Corollary \ref{PbndCor}.

\begin{thm} \label{NormalPbndCor}
Let $A$ be a normal matrix and  $(A, B)$ be a CIS input pair,
then the condition number of $P(A,B)$ satisfies the bound
\BEQ \label{Steinbound}
 \kappa_{}(P(A,B)) \ge \max \left\{
\frac{\lambda_n(A)^2 }{ \prod_{i=1}^n |\lambda_i(A)|^{2/d}}, \
\frac{\sigma_n\left( A'\right)^2 }{\lambda_n(A)^2 }, \
\frac{1 }{\lambda_1\left( A\right)^2 }
\right\} \ ,
\NEQ
where $\lambda_n(A)$ is the eigenvalue of $A$ with the smallest magnitude and
$A'$ is the IN matrix generated in the map defined in the proof of
Corollary \ref{PbndCor}. 
For $d=1$, the lower bound simplifies to
$\kappa_{}(P) \ge  1/ \prod_{i=1}^{n-1} |\lambda_i(A)|^2 $.
\end{thm}

We compare this bound to the condition number of $P(A, B)$
for an ensemble of input pairs where $A$ is a normal matrix; i.e.\ $A$
has orthogonal eigenvectors. We need to select a distribution
on the set of eigenvalue $n$-tuples. A natural choice is the distribution
$\nu_{\lambda}(\Lambda)$ induced by the random distribution of $A$ given
in Definition \ref{DefDAB1}.


\begin{defn} \label{NormDefDABN}
 $\Dc_N(A,B)$ is the set of CS input pairs, 
 $(\Lambda,B)$,  where
$\Lambda$  is diagonal. Let $\nu_{\lambda}(\Lambda,B)$
be the probability measure
induced from eigenvalue n-tuple distribution, $\nu_{\lambda,n}(\Lambda)$ of
$\nu_{n}(A,B)$ of Definition \ref{DefDAB1}
and  let $B_{ij}$ have the Gaussian distribution $N(0, 1)$ subject to the
controllability restriction.
\end{defn}

\begin{center}
\begin{tabular}{|c|c|c|c|c|c|c|c|} 
\hline
n  & 1\% & 10\%  & 25 \% & 50 \% & 75\%  & 90\% & 99 \% \\
\hline
8 & 7.27  & 9.25  & 10.8  & 12.4  & 14.6  & 16.7  & 21.7  \\
16& 15.0  & 17.7  & 19.4  & 21.3  & 23.6  & 26.1  & 31.6  \\
24& 23.2  & 25.9  & 27.7  & 29.8  & 32.1  & 34.5  & 39.3  \\
\hline
\end{tabular}
\end{center}

Table 2: Quantiles of $\log(\kappa)$ as function of $n$ for $d=1$.
{Note that $\log(\log(\kappa(P)))$ has an approximately
Gaussian distribution}

As seen in Table 2, our numerical computations show that
{\em the distribution of $\kappa(P)$ for the
normal matrices  $\Dc_N(A,B)$ is virtually identical to that of
our general random matrices $\Dc_N(A,B)$.}
Again, $\kappa(P)^{1/n}$ is approximately constant with median
condition number scaling as  $\kappa(P)^{1/n} \approx 3.4$.
The interquartile distance is again nearly independent of $n$
with a value of $\approx 4.4$.

\medskip
\begin{center}
\begin{tabular}{|c|c|c|c|c|c|c|c|} 
\hline
n  & 1\% & 10\%  & 25 \% & 50 \% & 75\%  & 90\% & 99 \% \\
\hline
8 & 1.52  & 2.68  & 3.54  & 4.83  & 6.55  & 8.60  & 13.4  \\
16& 2.19  & 3.45  & 4.51  & 5.86  & 7.57  & 9.72  & 14.8  \\
24& 2.83  & 3.97  & 4.90  & 6.27  & 8.1   & 10.1  & 15.3  \\
\hline
\end{tabular}
\end{center}

Table 3: Quantiles of $\log(\kappa / \kappa_{bd})$ as function of $n$.
Here $\kappa_{bd}= 1/ \prod_{i=1}^{n-1} |\lambda_i|^{2}$ is the bound
given in  Theorem \ref{Steinbound}, evaluated for each input pair.

Table 3 compares $\log(\kappa)$ versus our theoretical bound.
The discrepancy is growing only slightly in $n$, in contrast to
 $\log(\kappa)$ which is growing linearly in $n$. A regression
indicates that the median value of $\kappa / \kappa_{bd}$ is growing
as $n^{\alpha}$ with $1 \le \alpha \le 2$. Plotting the quantiles
of  $\log(\kappa/\kappa_{bd})$ as a function of $\log(\kappa_{bd})$
shows that the residual error is a weakly decreasing function of
 $\log(\kappa_{bd})$. We also observe that the spread of
        $\log(\kappa/\kappa_{bd} )$ is almost independent of
 of $\log(\kappa_{bd})$, perhaps indicating a heuristic model:
 $\log(\kappa) \sim \log(\kappa_{bd}) + f(n) + X_n $, where
the random variable $X_n$ barely depends on $n$. To model the
long tails of $\log(\kappa)$, an analogous model for $\log(\log(\kappa)))$
is probably called for.
We have also compared the normal bound with the log-condition number
for the ensemble of random matrices in Section \ref{GenSimSect}.
Surprisingly, the agreement with the bound is even better in this case.
However, there are many cases where the condition number of a random
input pair is smaller
than the bound for normal matrices predicts.

The bound (\ref{Steinbound}) indicates that $P$ can be quite ill-conditioned.
Theorems \ref{TransBnd} -\ref{NormalPbndCor} do not use any
property of $B$ (except controllability) nor of the complex phases of the
eigenvalues, $\lambda_i$.
Including this information in the bounds can only sharpen the lower bound.
We believe that a {\em significant fraction of the ill-conditioning
that is not explained  by our bound arises from using a random $B$.}
We could replace the quantile tables with analogous ones for
 $\inf_B \kappa(P(A,B))$. If we did, we would see that our bounds
better describe this quantity that the average value of $\kappa(P(A,B))$.

\ni
\section{ALTERNATING DIRECTION ITERATION BOUNDS}
\label{ADISect}

In this section we present condition number bounds based on the
alternating direction implicit (ADI) iteration for the solution of
the continuous time Lyapunov solution.
These results were formulated by T.\ Penzl in \cite{PenzlBnd}.
We restate his results in a more general context.

The results for the discrete Lyapunov equation 
follow by applying the bilinear transform. We define
$f(A, \tau) \equiv \left(A+ \tb {\mathbb{I}}_n  \right)^{-1}
        \left(A - \tau {\mathbb{I}}_n \right)$.
The Cayley transform corresponds to $\tau=1$: $\Ah = f(A, 1)$
and $\Bh = {\sqrt{2}}{\left({\mathbb{I}}_n + A\right)^{-1} B }$.
The solution $P(A,B)$ of the
discrete Lyapunov equation  (\ref{SteinEq}) 
for $(A,B)$ satisfies the Lyapunov equation
\BEQ \label{CALE}
\Ah P + P \Ah^*=- \Bh\Bh^* \ .
\NEQ
Following \cite{PenzlBnd}, we define the shifted ADI iteration on (\ref{CALE}).
To approximately solve (\ref{CALE}), we let 
 $P^{(0)} = 0$ and define $P^{(k)}$ by
\BEQ \label{ADIk}
P^{(k)} = f(\Ah, \tau_k) P^{(k-1)} f(\Ah, \tau_k)^* - 2 Re(\tau_k)
\left(\Ah+ \tb_k{\mathbb{I}}_n  \right)^{-1} \Bh\Bh^*
\left(\Ah^*+ \tau_k{\mathbb{I}}_n  \right)^{-1} \ ,
\NEQ
where the $\tau_k$ are the shift parameters. Using the methodology of \cite{PenzlBnd},
we have the following bound:

\begin{thm}
\label{ADIBndThm}
In the ADI iteration of (\ref{ADIk}), let $kd <n$.
The $P^{(k)}$ has rank $kd$
and satisfies the approximation bound:
\BEQ \label{ADIBnd}
\frac{\lambda_{kd+1}(P) }{\lambda_1(P)} \le
        \frac{\| P-P^{(k)} \|_2 }{\|P\|_2} \le
 \|F(\Ah;\tau_1 \ldots \tau_k)\|_2^2 \ ,
\NEQ
where $F(t; \tau_1,\ldots \tau_k) \equiv \prod_{j=1}^k f(t,\tau_j)$.
Let $\Ah$ have a complete set of eigenvectors and the eigenvalue
decomposition $\Ah = T \Lambda T^{-1}$,
then
\BEQ \label{ADIEval}
\|F(\Ah;\tau_1 \ldots \tau_k)\|_2^2 \le {\kappa}(T)^2
\max_{\lambda \in {\rm spec}({\Ah})} | F(\lambda, \tau_1,\ldots \tau_k) |^2 \ .
\NEQ
\end{thm}

We define
${\cal F}_k \equiv \min_{\tau_1,\ldots \tau_k}
\max_{\lambda \in {\rm spec}({\Ah})} |F(\lambda, \tau_1,\ldots \tau_k)|^2$.
Thus ${\cal F}_k$ is the best bound of the type in (\ref{ADIBnd})
{\em The difficulty in using Theorem \ref{ADIBndThm} is finding
good shifts that come close to approximating ${\cal F}_k$.}
There are algorithms for selecting shifts, but only rarely have
explicit upper bounds on ${\cal F}_k$ been given.
Penzl simplified this bound for the case of real, symmetric, stable $A$:



\begin{thm}[\cite{PenzlBnd}] \label{PenzlBndThm}
Let $\Ah$ be real symmetric and stable, and define
$\kh = \hat{\lambda}_1(A) / \hat{\lambda}_n(A)$. Then
\BEQ \label{PenzlBnd}
\frac{\lambda_{kd+1}(P) }{\lambda_1(P)} \le \left(
        \prod_{j=0}^{k-1}\frac{\kh^{(2j+1)/2k} -1} {\kh^{(2j+1)/2k} +1}
\right)^2\ .
\NEQ
\end{thm}

Penzl's proof is based on using a geometric sequence of shifts
on the interval containing the eigenvalues of $\Ah$.
It is difficult to determine when the bound (\ref{PenzlBnd}) is
stronger or weaker than the bounds in  Sections \ref{TransfSect}
and  \ref{FracSect} since  (\ref{PenzlBnd}) is independent of
the precise distribution of eigenvalues while (\ref{Steinbound})
uses the exact eigenvalues.

The bound (\ref{ADIEval}) shows that well-conditioned input pairs $(A,B)$
(such as input normal pairs) have $A$ and $\Ah$ that are
far from normal in the sense that {$\kappa(T)$} is large when
${\cal F}_k(A)$ is small.

\ni
\section{CONDITION BOUNDS FOR COMPANION MATRICES}
\label{CompSect}
We now specialize Corollary \ref{PbndCor} to the case where the advance matrix,
$A$, is a companion matrix. Other names for this case are
Frobenius normal form and Luenberger controller canonical form.
The second direct form \cite{MRbook} and
autoregressive (AR) models are  special case of this type and correspond
to $d=1$, with $B$ being the unit vector in the  first coordinate direction:
$B= {\bf e}_1$. For autoregressive models, $C= {\bf e}_1$, while the
second direct form uses $C$ to specify 
transfer function.
Let $A$ be of the form
\BEQ \label{CompForm}
A_c \equiv \left(
\begin{array}{cc}
-\mathbf{c}^{*} &-c_{0}^{*} \\
{\Pi}_{n-1} & \bf{ 0}
\end{array}
\right) \ ,
\NEQ
where ${\Pi}_{n-1} $ is a $(n-1) \times (n-1)$  projection matrix of the form
${\Pi}_{n-1}\equiv \Ib_{n-1}- \gamma {\bf e}_p {\bf e}_p^*$ where
$1 < p \le n-1$ and $\gamma=0$ or $1$.
Note that $\gamma=0$ corresponds to companion normal form.
Here $\mathbf{c}$ is an $(n-1)$ vector.

Autoregressive moving average (ARMA) models of degree $(p,q)$ 
satisfy the advance equation
$x_{t+1}+ c_1 x_t +c_2 x_{t-1}\ldots c_p x_{t+1-p} = 
e_{t+1} - c_{p+1} e_t +\ldots - c_{p+q} e_{t+1-q}$,
where $\{e_t\}$ is a sequence of independent random variables with
$E[e_t]=0$ and $E[e_t^2]=1$.
The ARMA $(p,q)$ model  has a state space representation with the
state vector
$z_t^{T} = (x_t, x_{t-1}\ldots x_{t-p+1}, e_{t}, \ldots e_{t-q+1})$,
 $B = {\bf e}_1 + \gamma {\bf e}_{p+1}$
 and $A$ given in (\ref{CompForm}) with $\gamma=1$ and $n=p+q$.
When $p=q$, this is a matrix representation of the first direct form.

\begin{lem} \label{CompSing}
Let $A_c$ be an $n \times n$ matrix of the form given
in (\ref{CompForm}) with $n > 2$,
then $A_c$ has singular values, $\sigma_1$ and $\sigma_m$, that
are the square roots of $\mu_{\pm}$,
where $\mu_{\pm}$ are the two roots of the equation
\begin{equation} \label{rootsComp}
\mu^{2}-\left( 1+\left| c_{0}\right| ^{2}+\left\| \mathbf{c}\right\|_{2}^{2}
\right) \mu+\left| c_{0}\right| ^{2} + \gamma \left| c_{p}\right| ^{2} =0 \ ,
\end{equation}
where $c_0$ and $\mathbf{c}$ are given in (\ref{CompForm}) and
$c_p$ is the $p$th component of the vector $\mathbf{c}$.
If $\gamma=1$, then $m=n-1$ and $A_c$ has a zero singular value.
Otherwise, $m=n$. 
The remaining singular values of $A_c$ are
one with multiplicity $n-2- \gamma$
and zero if $\gamma=1$.%
\end{lem}

For $\gamma=0$, this result is in \cite{Kittaneh}.
For $\gamma=1$, $\eb_{p+1}$ is a left null vector of $A_c$.

\Prf
Note $A_c A_c^{*} = \alpha \oplus {\Pi}_{n-1} - \wb {\mathbf{e}}_{1}^* -
\eb_{1}\wb^*$, where $\alpha \equiv \left| c_{0}\right| ^{2}+\left\| {\mathbf{c}}\right\|_{2}^{2}$, 
$\wb\equiv \left( \begin{array}{c} 0 \\ \Pi_{n-1} {\bf c} \end{array}\right)$.
To compute the eigenvalues of $A_c A_c^{*}$ we define an orthogonal
transformation to reduce $A_c A_c^{*}$ to the direct sum of
a $2\times 2$ matrix with  roots given by (\ref{rootsComp}) and
a projection matrix.
Let $H$ be the $(n-1) \times (n-1)$ Householder transformation such that
$H {\Pi}_{n-1}{\mathbf{c}}= \beta \eb_{p}$,
where $\beta = \|\Pi_{n-1} {\bf c} \|_2$, 
$\eb_{p}$ is the unit vector in $p$th coordinate.
We define  
and $\vb\equiv H{\eb}_p = \Pi_{n-1} {\bf c} / \beta$.
Since $H{\Pi}_{n-1}H^* = {\Ib}_{n-1}- \gamma H\eb_p {\bf e}_p^*H^*
 = {\Ib}_{n-1}- \gamma vv^*$, 
\begin{equation}\label{Comp2}
\left( \begin{array}{cc}
1 & 0 \\ 0 & H\end{array}\right)
A_c A_c^{*} \left( \begin{array}{cc}
1 & 0 \\ 0 & H\end{array}\right)^{*}=
\alpha \oplus \left( {\Ib}_{n-1}- \gamma \vb\vb^* \right) -\beta 
\left( {\mathbf{e}}_{p+1} {\mathbf{e}}_{1}^* + {\mathbf{e}}_{1} {\mathbf{e}}_{p+1}^*\right)
\end{equation}
If $\gamma=1$, then $\vb$
has a zero $p$th coordinate.
For both the $\gamma=0$ and the $\gamma=1$ cases,
 the eigenvalue equation decouples into
the eigenvalues of the  $2 \times 2$ matrix of the first and $(p+1)$st rows
and columns and the eigenvalues of ${\Ib}_{n-1}- \gamma vv^* $ projected onto
the space orthogonal to $\eb_p$.
The eigenvalue equation for the $2 \times 2$ matrix 
is given by 
$(\mu -\alpha)(\mu -1) - \beta^2=0$.
\eopp

We define
$\Gamma\equiv 1+\left| c_{0}\right| ^{2}+\left\| \mathbf{c}\right\|_{2}^{2}$
and $\omega = \left| c_{0}\right| ^{2}+ \gamma \left| c_{p}\right| ^{2}$.
We denote the largest root of (\ref{rootsComp}) by $\mu_{+}$ and 
the smallest by $\mu_{-}$. Note $\mu_{-} =\omega \mu_{+}^{-1}$.
The bound of Corollary \ref{PbndCor} reduces to

\begin{thm} \label{FrobPbndCor}
Let $A_c$ be  companion matrix as specified by (\ref{CompForm})
and  $(A_c, B)$ be a stable, controllable input pair with $n > 2$,
then the condition number of $P(A_c,B)$ satisfies the bound
$ \kappa_{}(P(A_c,B)) \ge  \mu_{+}$.
If $A_c$ is invertible,
\BEQ \label{FFbnd}
 \kappa_{}(P(A_c,B)) \ge \max \left\{ \mu_{+},
\frac{\sigma_n(A')^2 }{\mu_{-} }  \right\} \ ,
\NEQ
where $A'$ is the IN matrix
generated in the map defined in the proof of Corollary \ref{PbndCor}. 
Here $\mu_+$ satisfies
\begin{equation} \label{ContFrac}
1+\left\|\Pi_{n-1} \mathbf{c}\right\| _{2}^{2} \ \le \
\Gamma-\frac{\omega}{\Gamma -{\omega} }
\le \ \mu_{+}
\le \
\Gamma-\frac{\omega}{\Gamma -\frac{\omega}{\Gamma}} \le \Gamma.\
\NEQ 
\end{thm}

\Prf
The bound (\ref{ContFrac}) is proven by 
rewriting (\ref{rootsComp}) as a sequence of continued fractions
\BEQ\label{rootsRewrite}
\mu = \Gamma -\frac{\omega}{\mu }
=
\Gamma-\frac{\omega}
{\Gamma -\frac{\omega}{\mu }} \ . 
\NEQ
Applying simple bounds to the continued fractions yields (\ref{ContFrac}).
\eopp

The bound in Theorem \ref{FrobPbndCor} also applies
when $A^*$ is in companion form,
corresponding to Luenberger observer canonical form.
If the eigenvalues of $A_c$ are prescribed, 
the coefficients in $A_c$, $\{c_k\}$,
are the coefficients of the characteristic polynomial of $A_c$:
$p(\lambda)= \prod_{i=1}^n (\lambda-\lambda_i) =
\lambda^{n}-\sum_{i=1}^{n-} c_i \lambda^{n-i}- c_0$ for $\gamma=0$.
When  the eigenvalues of $A_c$ are positive real,
a weaker but explicit bound is

\begin{thm} \label{FrobPos}
Let $A_c$ be invertible with positive real eigenvalues $\lambda_i$,
then $\kappa(P(A_c,B))$ $ >  \prod_{i=1}^n (1 +|\lambda_i|)^2 /(n+1) - \prod_{i=1}^n|\lambda_i|^2$.
\end{thm}

\Prf
We evaluate the characteristic polynomial at $-1$:
$|p(-1)| = \prod_{i=1}^n (1 +|\lambda_i|)= \sum_{j=0}^n |c_j| $,
where  $c_n\equiv1$. 
Note $(\sum_{j=0}^n |c_j|)^2 /(n+1) < \sum_{j=0}^n |c_j|^2$ and
$|c_0|^2 =\prod_{i=1}^n|\lambda_i|^2$.
The bound (\ref{ContFrac}) implies
$\mu \ge \sum_{j=1}^n |c_j|^2 \ge  (\sum_{j=0}^n |c_j|)^2 /(n+1) -|c_0|^2$.
\eopp

For $n>>1$, $|c_0|^2 << \prod_{i=1}^n (1 +|\lambda_i|)^2 /(n+1)$.
When the eigenvalues of $A_c$ have random or near random complex phases,
the value of $\mu_+$ is typically much less than
 $\prod_{i=1}^n (1 +|\lambda_i|)^2 /n$,
since generally $\sum_i \lambda_i << \sum_i |\lambda_i|$,
 $\sum_{i \ne j} \lambda_i \lambda_j << \sum_{i\ne j} |\lambda_i\lambda_j|$,
 $\ldots$.
We now compare the bound in Theorem \ref{FrobPbndCor}
with a random distribution of $A_c$.






\begin{defn} \label{CombDefDAB1}
$\Dc_C(A_c,B)$ is the set of stable, controllable $(A_c, B)$,
where $A_c$ is given by (\ref{CompForm}) and $B= \eb_1$.
The distribution $\nu_C(A,B)$ is defined by the eigenvalues of $A_c$
having the distribution $\nu_{\lambda,n}(\Lambda)$ of
$\nu_{n}(A,B)$ of Definition \ref{DefDAB1}
subject to the CS restriction.
\end{defn}

Table 4a gives the quantiles of $\log(\kappa(P))$ for $A_c$ with random $B$
and Table 4b gives the corresponding quantiles for $B=\eb_1$.
The random $B$ case has a very broad distribution with the
interquartile distance two to three times larger than that of
generic random $(A,B)$ case. The top quartile of the random $B$
Frobenius case is as badly conditioned as the random case although
the bottom quartile is much better conditioned.

\medskip

\begin{center}
\begin{tabular}{|c|c|c|c|c|c|c|c|} 
\hline
n  & 1\% & 10\%  & 25 \% & 50 \% & 75\%  & 90\% & 99 \% \\
\hline
8 & 2.30  & 4.08  & 5.66  & 8.68  & 12.2  & 15.1  & 21.1  \\
16& 4.10  & 6.71  & 9.59  & 14.6  & 20.1  & 24.2  & 29.6  \\
24& 5.61  & 8.57  & 12.4  & 19.3  & 27.5  & 32.5  & 38.7 \\
\hline
\end{tabular}
\end{center}

Table 4a: 
Quantiles of $\log(\kappa(P))$ distribution for $\Dc_C(A_c,B)$ for randomly
distributed $B$. 
The bound \ref{FFbnd} significantly underestimates
 $\kappa(P)$ in many cases.

We have also computed the condition numbers when the eigenvalues are
all positive: $ \lambda_i \rightarrow |\lambda_i|$.
These models are appreciably more ill-conditioned than 
in cases where the eigenvalues are distributed randomly in the complex plane.
{\em This ill-conditioning corresponds to the difficulty in estimating
the coefficients of a sum of decaying exponentials.}
For positive $\{\lambda_i\}$, the observed ill-conditioning is usually
much larger than the
formula, $\kappa >  \prod_{i=1}^n (1 +|\lambda_i|)^2 /(n+1) -\prod_{i=1}^n |\lambda_i|^2$.

The $B=\eb_1$ case  may be interpreted as a random autoregressive model.
Our distribution, $\Dc_C(A_c,B)$, corresponds to a random distribution of
poles of the autoregressive transfer function:
For autoregressive models, the observability Grammian corresponds to solving
(\ref{SteinEq}) with the pair $(A^*, A[1,:]^t)$, i.e.\ using 
the characteristic polynomial coefficients as $B$.
Thus we may examine the condition numbers of both the controllability
and observability Grammians.

\medskip
\begin{center}
\begin{tabular}{|c|c|c|c|c|c|c|c|} 
\hline
n  & 1\% & 10\%  & 25 \% & 50 \% & 75\%  & 90\% & 99 \% \\
\hline
8 & .825  & 1.69  & 2.45  & 3.59  & 5.05  & 6.62  & 9.89  \\
16& 1.66  & 2.88  & 3.79  & 5.08  & 6.74  & 8.48  & 11.8  \\
24& 2.36  & 3.66  & 4.64  & 5.99  & 7.66  & 9.36  & 13.1  \\
\hline
\end{tabular}
\end{center}

Table 4b: 
Quantiles of $\log(\kappa(P))$ distribution for $\Dc_C(A_c,B)$ for $B=\eb_1$.

The autoregressive models ($B=\eb_1$) are much better
conditioned than those with a random righthand rank-one side.
The scaling of the median of $\log(\kappa(P))$ versus $n$ is ambiguous.
The interquartile distance of $\log(\kappa(P))$ is a weak function of $n$.
Table 4c examines the observability Grammian of the autoregressive model.
We find that the corresponding observability Grammians
for our autoregressive models are very poorly conditioned. 
Thus these autoregressive models are nearly unobservable.

\medskip
\begin{center}
\begin{tabular}{|c|c|c|c|c|c|c|c|} 
\hline
n  & 1\% & 10\%  & 25 \% & 50 \% & 75\%  & 90\% & 99 \% \\
\hline
8 &  14.8  & 22.1  & 27.2  & 35.1  & 46.8  & 61.5  & 89.3  \\
16 & 42.2  & 53.9  & 64.4  & 79.6  & 95.5  & 101  & 109  \\
24 & 72.5  & 92.1  & 101  & 104  & 108  & 111  & 117  \\
\hline
\end{tabular}
\end{center}

Table 4c: 
Quantiles of $\log(\kappa(P))$ for the observability Grammian of the autoregressive model.


The controllability is much better conditioned for the random autoregressive 
model than for the normal advance matrices with the same spectrum
while the observability Grammian is grossly ill-conditioned in
the random autoregressive model.  These results 
may be influenced by the choice of $\Dc_C(A_c,B)$.

\ni
\section{POWER ESTIMATE}
\label{PowerSect}
We now show that for certain classes of input pairs, the condition number,
$\kappa \left( P(A,B) \right)$, grows exponentially in $n/d$. Specifically,
let $\{ (A_n,B_n) \}$ be CS  with the uniform bound $\sigma_1(A_n) \le c<1$.
The proof applies Lemma \ref{AbstractBnd} with  $\phi(A)= A^k$.
The theorem below shows that
$\kappa \left( P(A_n,B_n) \right) \ge c^{-2 k}$,
where $k$ is the integral part of $(n-1)/d$.

\begin{thm} \label{PowThm}
Let $(A, B)$ be a stable, controllable input pair,   then
$\kappa \left( P(A,B) \right) \ge \left| \sigma _{1}\left( A\right) \right|^{-2k}$,
where $kd<n$.
\end{thm}

\Prf
Let $T$ be a INizing transformation, $A' = TAT^{-1}$, $B'= TB$ with $(A',B')$ being IN.
Let $\phi(A)= A^k$ for $k$ such that
$kd<n$. By Lemma \ref{AbstractBnd},
$\kappa \left( P(A,B) \right) \ge
\left| \sigma _{1}\left( A'^k\right) \right|^2 / \left| \sigma _{1}\left( A^k\right) \right|^2$.
Note $\left| \sigma _{1}\left( A^k\right) \right| <
        \left| \sigma _{1}\left( A\right) \right|^k$.
The proof is completed when we prove the lemma below that
$\left| \sigma _{1}\left( A'^k\right) \right| =1$.
\eopp

\begin{lem} \label{PowLem}
Let $(A, B)$ be a IN pair, then $\sigma_{1}\left( A^{k}\right) =1$ for  $kd<n$.
\end{lem}

\Prf
Let $k$ be such that $kd<n$, then
$\limfunc{rank}\left( B | AB | \cdots | A^{k-1}B\right) \le kd$ and
\begin{equation}
A^{k}A^{k*}= {\mathbb{I}}- \left(
\begin{array}{cccc}
B & AB & \cdots & A^{k-1}B
\end{array}
\right) \left(
\begin{array}{cccc}
B & AB & \cdots & A^{k-1}B
\end{array}
\right) ^{*} \ .
\end{equation}
Thus
$\sigma _{1}\left( A^{k}\right) =1$
for $kd<n$.
\eopp

In particular, when $A$  is normal and $d=1$, $k=n-1$, the bound becomes
 $\kappa \left( P\right) \ge  \left| \lambda _{1} \right|^{- 2 (n-1) }$.

\ni
\section{JORDAN BLOCK BOUNDS} \label{JordanSect}
We now bound $\kappa_{}(P(A,B))$ when $A$ is a single Jordan block:
$A = J_o\equiv \lambda_o {\mathbb{I}}_n + Z_n$,
where $Z_n$ is the lower shift matrix:
$Z_{i,j} = \delta_{i-1-j}$.
Our bound shows that for $d=1$, the condition number grows as
$ (1-|\lambda_o|)^{2-2n}$ when $n>>1$.
The proof takes the maximum of the bounds  for
 $\phi(A)= A^k$ based on  Lemma \ref{AbstractBnd}

\begin{thm} \label{JordanBoundthm}
Let $(J_o, B)$ be a CS input pair with
$J_o=\lambda_o {\mathbb{I}}_n + Z_n$ with $0< | \lambda_o|<1$ and $d=1$.
The condition number of $P(J_o,B)$ satisfies the bound
\BEQ \label{Jordbound}
\kappa_{}(P(J_o,B)) \ge \left[\frac{1}{n}
\left( \frac{1-\frac{1}{n}}{1-|\lambda_o|}\right)^{n-1} \right]^2 >
\left[ \frac{e^{-1}}{n }\left( \frac{1}{1-|\lambda_o| }\right)^{n-1} \right]^2 \ .
\end{equation}
\end{thm}

Prior to proving Theorem \ref{JordanBoundthm}, we state the bounds for
$\sigma_n$.
We define $J(\lambda)=  \lambda {\mathbb{I}}_n + Z_n$.

\begin{lem} \label{JordanSVD}
Let $J(\lambda) = \lambda {\mathbb{I}}_n + Z_n$ with $\lambda \ne 0$, then
$\sigma_n(J(\lambda)) \le  |\lambda|^n$.
\end{lem}

The bound is well-known and follows from
$\left[ J\left( \lambda \right) +(-1)^{n}\lambda^{n}\eb_{1}\eb_{n}^{T}\right] v= 0$
where $v_j = (-1)^j \lambda ^{n-j}$. \eopp

{\em Proof of Theorem \ref{JordanBoundthm}:}
Let $T$ be the transformation  from $(J_o,B)$ to an IN pair, $(A',B')$.
By Lemma \ref{PowLem},
$\sigma_1(A'^k) \le \sigma_1(A')^k \le 1$ for $k>0$.
From Lemma \ref{AbstractBnd},
$\kappa_{}(T) \ge \sup_{k\ge 0}\left( \sigma _{1}\left( J_o^{k}\right) /  \sigma_1(A'^k) \right) \ge
\sup_{k\ge 0} \sigma _{1}\left( J_o^{k}\right)$.
We now apply
the following  version of the Kreiss Matrix Theorem:

\begin{lem}  \label{WTlemma}  \cite{WT}
Let $H$ be a stable $n \times n$ matrix, $\sigma _{1}\left( H\right) \ge 1$, then
\begin{equation}
\sup_{k\ge 0}\sigma _{1}\left( H^{k}\right) \ge \phi \left( H\right) \equiv
 \sup_{\left| z\right| >1}\frac{|z| -1 }{\sigma_{n}\left( z{\mathbb{I}}-H\right) }.
\end{equation}
\end{lem}
Setting $H=J_o$ yields 
\BEQ
\kappa_{}(T)\ge \phi \left( J_o \right) \ge
\frac{| z_o|-1  }
{\sigma _{n}\left(  z_o{{\mathbb{I}}}-J_o\right)}
\ge
\frac{ | z_o|-1  }{| z_o-\lambda_o|^n}
\NEQ
for any $z_o \ne \lambda_o$ using the bound on $\sigma _{n}$ from Lemma \ref{JordanSVD}.
Maximizing the expression in $z_o$ yields 
$ (\zb_o -\lb_o) z_o = n |z_o|(|z_o|-1)$. Solving yields the choice 
$|z_o| \equiv \frac{n- |\lambda_o| }{n-1}$ with the complex phase of $z_o$ equal to the phase of $\lambda$.
Inserting this value of $z_o$ yields (\ref{Jordbound}).
\eopp




Table 5 gives the quantiles of $\log(\kappa(P(J_o(\lambda),B)))$
averaged over the ensemble of $B$. We also give  the minimum value
of $\kappa$ that we observed over 1200 realizations and the bound.
We observe that the bound on the minimum value of $\kappa$ becomes
more accurate on for as $\lambda$ gets closer to one.
This seems reasonable since the bound in  (\ref{Jordbound}) grows
strongly as $|\lambda | \rightarrow 1$.
The interquartile distance appears to be growing as
 $IQ(\log(\kappa)) \sim g(\lambda) n$ indicating a wide spread.
Thus the typical value of $\kappa$ will be a factor of $\exp(ng(\lambda))$
larger than our bound.

\medskip
\begin{center}
\begin{tabular}{|c|c|c|c|c|c|c|}
\hline
n  & $\lambda$ & Bound & $\min$  & 10 \% & 50 \%   & 90\%  \\
\hline
  8  & 0.3  &  -1.17 &  3.49  & 6.03  & 11.6  & 32.9  \\
 16  & 0.3  &  3.16  &  9.80  & 14.8  & 23.7  & 66.1  \\
 24  & 0.3  &  8.05  &  19.5  & 23.7  & 36.3  & 89.7  \\
\hline
 8   & 0.5  &  3.55  &  7.23  & 9.49  & 14.0  & 32.5  \\
 16  & 0.5  &  13.2  &  18.6  & 23.8  & 31.2  & 68.1  \\
 24  & 0.5  &  23.5  &  32.1  & 38.2  & 47.6  & 94.4  \\
\hline
  8  & 0.8  &  16.4  &  19.2  & 21.0  & 23.9  & 34.7  \\
 16  & 0.8  &  40.7  &  44.4  & 48.1  & 51.8  & 76.4  \\
 24  & 0.8  &  65.7  &  70.2  & 75.7  & 80.7  & 95.7  \\
\hline
\end{tabular}
\end{center}

\medskip

Table 5:
Summary of $\log(\kappa(P))$ distribution over an ensemble of $B$
for fixed $\lambda$ and $n$.
Here $\kappa_{bd}$ is the bound given in  (\ref{Jordbound}).

\noindent
\section{BILINEAR TRANSFORMATIONS BOUNDS} \label{FracSect}

We now show that if $A$ is normal and all of the the eigenvalues
of $A$ are approximately equal ($\lambda_k \approx x$) then
 $\kappa(P)$ is exponentially large in $n/d$.
Our analysis is based on applying Theorem \ref{AbstractBnd}
to a fractional linear transformation of $A$.

We define the bilinear  map of $A$:
$ A \rightarrow \Ah \equiv f(A,w)$, $ B \rightarrow \Bh\equiv g(B,w)$ defined by
\BEQ \label{BiMapDef}
\Ah = f(A, w) \equiv \left({\mathbb{I}}_n - w^* A\right)^{-1}
        \left(A -w {\mathbb{I}}_n \right), \ \
\Bh = {\sqrt{1-|w|^2}}{\left({\mathbb{I}}_n - w^* A\right)^{-1} B }
\NEQ
with $|w|<1$.
The bilinear map, $f(z, w)= (z-w)/ (1- w^* z)$,
is a univalent function that maps the unit disk $|z|<1$ onto itself and
thus preserves stability.
The bilinear map preserves solutions of the (\ref{SteinEq})
$P(\hat{A},\hat{B}) = P(A,B) $.
Let $T$ be an INizing transformation of $(A,B)$ to  $(A', B')$.
Note $f(TAT^{-1},w) = Tf(A,w)T^{-1}$ so $T$ is also
an INizing transformation of $(\hat{A},\hat{B})$
to the IN pair $(f(A',w), g(B',w))$.
We now bound the condition number of $P=T^{-1}T^{-*}$ by applying
Theorem \ref{TransBnd} to $\hat{A}=f(A;w)$.
Note $\hat{A}$ has eigenvalues, $\{ f(\lambda_i(A),w)\}$
and is normal if $A$ is.
Thus $\sigma_n(f(A,w)') \le \prod_{i=1}^n |f(\lambda_i(A),w)|^{2/d}$.

\begin{thm} \label{FracNormalThm}
Let $A$ be a normal matrix and  $(A, B)$ be a CS input pair
with nonsingular $f(A,w)$ and $|w|<1$,
then the condition number of $P(A,B)$ satisfies the bound
\BEQ \label{FracSteinbound}
 \kappa_{}(P(A,B)) \ge
\frac{\min_k\{|f(\lambda_k,w)|^2\}}{ \prod_{i=1}^n |f(\lambda_i,w)|^{2/d}} \ .
\NEQ
\end{thm}

Theorem \ref{FracNormalThm} allows us to optimize $w$ in the bound
(\ref{FracSteinbound}) for a given set of eigenvalues.
Theorem \ref{NormalPbndCor} corresponds to $w=0$.
As $w$ tends to zero, we see that Theorem \ref{NormalPbndCor}
remains valid when $A$ is singular.


We now assume that the eigenvalues are localized in a
shifted disk: $|\lambda_i - x| < \rho$ where $|x| + \rho < 1$.
Here $x$ is the center of the disk which contains all of the eigenvalues
and $\rho$  is the radius.
We now return to the ansatz that all of the eigenvalues of $A$ are contained
in the disk of radius $\rho$ centered about $x$.
Choosing $w=x$, the bilinear transformation maps the circle,
$z(\theta) = x + \rho e^{i\theta}$, to the circle
 $|\lambda|< \rho /(1-|x|\rho -\rho^2)$.
Applying Theorem \ref{PowThm} yields

\begin{thm} \label{DiskBoundthm}
Let $A$ be a normal matrix and  $(A, B)$ be a stable, controllable input pair
with the eigenvalues are localized in a
shifted disk: $|\lambda_i - x| < \rho$, where $|x| + \rho < 1$.
Let $kd< n$.  
The condition number of $P(A,B)$ satisfies the bound
$ \kappa_{}(P(A,B)) \ge \left( \frac{\rho}{1-|x|\rho - \rho^2}\right)^{-2k}$.
\end{thm}

 This bound illustrates the ill-conditioning that results when the eigenvalues
of $A$ are clustered in the complex plane.

\section{COLORED NOISE FORCING}
\label{ColSect} 

In \cite{NH}, a computation is presented that shows that input
normal filters perform well in the presence of autocorrelated noise.
We now examine the condition number of the controllability Grammian
when forcing term is autocorrelated. Our results help to explain the
good performance of IN pairs observed in \cite{NH}.
Let the state vector, $z_t$, evolve as
\BEQ \label{corAdv}
z_{t+1} =Az_{t}+Bx_{t} =Bx_{t}+ABx_{t-1} + A^2Bx_{t-2}+ \cdots \ ,
\NEQ
where $x_t$ is a zero mean stationary sequence with the $ d \times d$
autocovariance, $\phi _{k}=E\left[ x_{t}x_{t-k}^{*}\right] $.
The covariance of the state vector, $ W\equiv E\left[ z_{t}z_{t}^{*}\right]$,
satisfies
\begin{equation} \label{zzColor}
W =\left(
\begin{array}{cccc}
B & AB & A^{2}B & \cdots
\end{array}
\right) \left(
\begin{array}{lll}
E\left[ x_{t-1}x_{t-1}^{*}\right]  & E\left[ x_{t-2}x_{t-1}^{*}\right]  &
\\
E\left[ x_{t-1}x_{t-2}^{*}\right]  & E\left[ x_{t-2}x_{t-2}^{*}\right]  &
\ddots  \\
& \ddots  & \ddots
\end{array}
\right) \left(
\begin{array}{c}
B^{*} \\
B^{*}A^{*} \\
B^{*}A^{*2} \\
\vdots
\end{array}
\right) \ .
\end{equation}

\begin{thm}
Let $\left( A,B\right) $ be a CS input pair and $x_{t}$ a zero mean
stationary sequence with autocovariance $\Phi $, $\Phi _{jk}=E\left[
x_{t-k}x_{t-j}^{*}\right] $. Let $z_{t}$ be a sequence of state vectors
satisfying (\ref{corAdv}),
where $x_t$ is a stationary sequence with a smooth spectral density.
Let $W=E\left[ z_{t}z_{t}^{*}\right] $ be the state covariance. Then
\begin{equation}
\kappa \left( W\right) \le \kappa \left( P\right) S_{\max }/S_{\min } \ ,
\end{equation}
where $S_{\min }$ and $S_{\max }$ are the minimum and maximum modulus of the
spectral density of $x_{t}$ and $P$ is the solution of $P-APA^{*}=BB^{*}$.
\end{thm}

Pessimists (and many realists)  expect that 
$\kappa \left( W\right) \sim \kappa \left( P\right) S_{\max }/S_{\min }$.

\emph{Proof:}\thinspace . Let $M_{t}\equiv (B\ AB\ A^{2}B\ \ldots A^{t-1}B)$
and let $\Phi _{t}$ be the $dt\times dt$ leading submatrix of the noise
covariance matrix $\Phi $. Define $W_{t}\equiv M_{t}\Phi _{t}M_{t}^{*}$, by
the lemma below we have
\begin{equation}
\kappa \left( W_{t}\right) \le \kappa \left( M_{t}\right) ^{2}\kappa \left(
\Phi _{t}\right) \   \label{ColRes}
\end{equation}
and by well-known result that $\kappa \left( \Phi _{t}\right) \le {S_{\max }}%
/{S_{\min }}$ \cite[p.\ 137]{BD}
\begin{equation}
\kappa \left( W_{t}\right) \le \kappa \left( M_{t}\right) ^{2}{S_{\max }}/{%
S_{\min }.}
\end{equation}
{\ Let }$M\equiv (B\ AB\ A^{2}B\ \ldots )$; then $MM^{*}-AMM^{*}A^{*}=BB^{*}$%
, and $MM^{*}=P$. Since $A$ is stable, $M_{t}M_{t}^{*}=\sum_{k=0}^{t-1}
A^{k}BB^{*}\left(A^*\right)^k$ converges to $P$. Similarly $W_{t}$
converges to $W$ as $t$ increases. Singular values are continuous functions
of the matrices $M_{t}$ and $W_{t}$ so the result follows by taking the
limit on both sides. {\vrule height7pt width7pt depth0pt}

We now prove (\ref{ColRes}):

\begin{lem}
Let $\Phi $ be Hermitian positive definite and $M$ be a $r\times n$ matrix
of rank $r$, then $M\Phi M^{*}$ satisfies
$\kappa \left( M\Phi M^{*}\right) \le \kappa \left( \Phi \right) \kappa
\left( M\right) ^{2}$ . 
\end{lem}

\noindent \emph{Proof:}\thinspace  Clearly $\sigma _{1}\left( M\Phi
M^{*}\right) \le \sigma _{1}\left( M\right) ^{2}\sigma _{1}\left( \Phi
\right) $. 
Note
\begin{equation}
\sigma _{r}\left( M\Phi M^{*}\right) =\min_{v}\Vert M\Phi M^{*}v\Vert =
\min_{v}|v^{*}M\Phi M^{*}v|\ge \sigma _{r}\left( M\right) ^{2}\sigma
_{n}\left( \Phi \right) \ ,
\end{equation}
where $v$ has norm one. Dividing the first inequality by the second proves
yields the lemma. {\ \vrule height7pt width7pt depth0pt}

For input normal realizations, the bound is $\kappa \left( W\right) \le
\kappa \left( \Phi \right) \le \frac{S_{\max }}{S_{\min }}$.
Colored noise forcing arises when a signal is being undermodeled or
modeled with uncertainty. 
The bound 
shows that input normal representations
have state covariances that are well-conditioned even in the presence
of colored noise. 
This property is independent of the system order and is important
for practical applications.
The lower bounds given in previous sections show that many
common or random state space representations can be expected to fail for
high order systems even for white noise.

\ni
\section{SUMMARY} \label{V}

We have examined the condition number, $\kappa(P)$, of solutions of
the discrete Lyapunov equation. 
For random stable controllable input pairs, the $n$th root of the
condition number, $\kappa^{1/n}(P)$, is approximately constant.
When $A$ is normal with the same distribution of eigenvalues,
 $\kappa(P)$ has a very similar distribution.
In both these cases, the median condition number grows exponentially
while the interquartile distance of $\log(\kappa)$ has a weak dependence
on $n$. Empirically, $\log(\log(\kappa(P)))$ has an approximately
Gaussian distribution.

We have given analytic bounds for the conditioning of solutions
of the discrete Lyapunov equation. 
For cases with $n>>d$, these bounds can be
considerable.  For both normal advance matrices
and random advance matrices, the analytic bound for normal $A$ explains
a large portion of the ill-conditioning. Nevertheless, the actual condition
numbers  are often several hundred times larger or more. The ill-conditioning,
and the excess ill-conditioning, $\kappa(P)/\kappa_{bd}$, are larger
when the eigenvalues cluster in the complex plane, (either as a single
Jordan block or as multiple closely spaced eigenvalues).


For random autoregressive models,
the controllability Grammian is usually well-conditioned and the observability
Grammian is extremely ill-conditioned (for our ensemble of models).



Our analytic bounds do not use any property of $B$ except controllability.
Thus our results are actually lower bounds on $\inf_B \kappa(P(A,B))$.
Our bounds in Sections IV-VII do not utilize information on
the complex phases of the eigenvalues, $\lambda_i$.
Including additional information in the bounds
can only sharpen the lower bound. Alternatively, we could compare our bounds
versus the best possible $B$ matrix for a given $A$.

Finally, we have examined the covariance of the state vector in the presence
of autocorrelated noise. Our bound depends on the ratio of the maximum to the minimum
spectral density of the noise. When this ratio is not to large and
an input normal representation is used,
then the covariance of the state vector is well-conditioned.
This indicates that input normal representations are robust to undermodeling
errors in filter design and system identification.



\end{document}